\documentclass[twocolumn,eqsecnum,superscriptaddress,showpacs]{revtex4}
\usepackage{graphics}
\usepackage{latexsym}
\usepackage{bm}
\begin{document}
\title{Novel Solution of Mercury Perihelion Shift}

\author{Takehisa Fujita}\email{fffujita@phys.cst.nihon-u.ac.jp}
\author{Naohiro Kanda}\email{nkanda@phys.cst.nihon-u.ac.jp}
\affiliation{Department of Physics, Faculty of Science and Technology, 
Nihon University, Tokyo, Japan}

\date{\today}%

\begin{abstract}

We present a novel solution of the Mercury perihelion advance shift in the new 
gravity model. It is found that the non-relativistic reduction of the Dirac 
equation with the gravitational potential produces the new gravitational 
potential of $\displaystyle{ V(r)=-{GMm\over r}+{G^2M^2m^2\over 2mc^2r^2 } }$. 
This potential can explain the Mercury perihelion advance shift without any free 
parameters. Also, it can  give rise to the $\omega-$shift of the GPS satellite where 
the advance shift amounts to $\left({\Delta \omega\over \omega}\right)_{th}
 \simeq 3.4\times 10^{-10}$ which should be compared to the recent observed value of  
$\left({\Delta \omega\over \omega}\right)_{exp} \simeq 4.5\times 10^{-10}$.

\end{abstract}


\maketitle

\noindent

\section{Introduction}

In a recent paper, a new gravitational model is proposed which can well describe  
all the known properties of the gravity physics \cite{tfgrav,fujita}. 
In addition, this model predicts a new interaction between photon and gravity 
at the fourth order Feynman diagram, and this photon gravity interaction should 
play some important role in astrophysics. 

From this gravity model, we obtain, for the first time, a proper Dirac equation 
with the gravitational potential, and it is written as 
$$ \left[-i \bm{\nabla}\cdot \bm{\alpha}+ \left(m -{GmM\over r}\right) 
\beta \right] \psi =E\psi  \eqno{(1.1)} $$
where $G$ and $M$ denote the gravitational constant and the mass of the gravity center, 
respectively. This equation can be reduced to the non-relativistic equation in terms of 
Foldy-Wouthuysen transformation, and after we make the classical limit, the new 
gravitational potential becomes
$$ V(r)=-{GMm\over r}+{G^2M^2m\over 2c^2}{1\over r^2 }.  \eqno{(1.2)} $$
This additional potential is special for the scalar potential, 
in contrast to the Coulomb potential which has no such effect from 
the non-relativistic reduction of the Dirac equation. 

In this paper, we show that the new gravitational potential can explain 
the advance shift of the Mercury perihelion to a remarkably good accuracy. 
Also, it gives rise to the advance $\omega-$shift of the GPS satellite 
$ \left({\Delta \omega\over \omega}\right)_{th} \simeq  3.4 \times 10^{-10}  $ 
which should be 
compared to the observed value of ${ \left({\Delta \omega\over \omega}\right)_{exp} 
\simeq  4.5 \times 10^{-10} }$ \cite{bah}, and one sees 
that the theoretical evaluation achieves a remarkable agreement with experiment. 
Surprisingly, all the calculations are carried 
out without any free parameters, and this strongly suggests that 
the new gravity model must be a correct theory of gravitation. 

In addition, this model is applied to the description of the advance shift 
of the earth rotation around sun \cite{leap}. The predicted time shift 
for the earth rotation around the sun for one year is calculated to be
$$ (\Delta T)_{th} \simeq  0.621  \ \ s/year      $$
which should be compared to the observed time shift in terms of the leap second
$$ (\Delta T)_{exp}  \simeq  0.63 \pm 0.02 \ \ s/year  .  $$
The agreement between the prediction and the observation is perfect, and this is 
discussed in detail in \cite{leap}.

It should be fair to make a comment on the general relativity \cite{ein}. 
However, the intrinsic problems of the general relativity are well described and 
discussed in detail in \cite{tfgrav,fujita}. Therefore, we discuss only 
the numerical results of the wrong direction of the Mercury perihelion shift 
by the general relativity. The basic mistake arises from the calculation of the angular 
shift instead of the angular velocity shift. This is quite clear since the angular shift 
is not a physical observable. 
To understand intuitively a simple reason for the advance or retreat shifts of 
the Mercury perihelion, we discuss the shift in terms of the physical 
observable $\omega T$ with $\omega$ angular velocity and $T$ the period. 
This quantity $\omega T$ is basically proportional to 
the elliptic area of the Mercury orbit, and the new additional 
potential is repulsive and therefore it gives a bit larger area than the Newton solution, 
while the general relativity is attractive, and therefore it gives a smaller area. 
This indeed corresponds to the advance or retreat shifts of the Mercury perihelion, 
and we will show it later more in detail.

\section{New Gravity Model }
In this section, we briefly review the new gravity model \cite{tfgrav,fujita}. 
First, we write the Lagrangian density 
$$  {\cal L} =  i\bar \psi  \gamma^{\mu}{\partial}_{\mu}\psi -e\bar \psi\gamma^{\mu}
 A_{\mu}\psi -m(1+g{\cal G}) \bar \psi \psi $$
$$  -{1\over 4} F_{\mu\nu} F^{\mu\nu}  
+ {1\over 2}\partial_\mu {\cal G} \ \partial^\mu {\cal G}  \eqno{(2.1)} $$
where $A_{\mu}$ and ${\cal G}$ denote the electromagnetic field and gravitational 
field, respectively. The gravity is a massless scalar field which couples 
to the fermion field $\psi $ in the mass term. This Lagrangian density is 
gauge invariant, and therefore it keeps the most important local symmetry. 
 
\subsection{Photon-Gravity Interaction}
As shown in \cite{tfgrav,fujita}, photon interacts with gravitational field 
at the fourth order of Feynman diagram. The detailed expression is given in 
\cite{tfgrav,fujita}, and therefore we write only the result of the equation of motion 
for photon $\bm{A}(t,\bm{r}) $ under the gravitational field
$$ \left( {\partial^2\over{\partial t^2}}-\bm{\nabla}^2 -{ G\alpha m^2_t M\over 2\pi}
{1\over r} \right)\bm{A}(t,\bm{r}) =0   $$
where $M$, $m_t$ and $\alpha$ denote the mass of the gravitational center, the total 
mass of different kind of fermions and the fine structure constant, respectively. 
An interesting question may be as to whether there is some experiment which can measure 
the photon gravity interaction.  We believe that, if one makes use of the satellites, 
there may well be some chance to measure it because of the laser technique. From one 
satellite to the other, one can shoot some laser beam with very low energy. In this 
case,  this laser beam may well be scattered by the earth gravity, and the scattered 
laser may be observed at the other satellite detector. 

\subsection{Proper Dirac Equation in Gravitational Potential}
For a long time, the understanding of the Dirac equation in the gravitational 
potential has been incomplete. That is, a proper Dirac equation for fermions 
was not known in the presence of the gravitational potential. 
This was not a healthy situation for theoretical physics since, it indicates that, 
even the Newton equation with the gravitational field cannot be derived properly. 
It is, in this respect, quite important that we have now the Dirac equation 
for fermions in the gravitational potential. 
From the Lagrange equation of eq.(2.1), we obtain
$$    i \gamma^{\mu}{\partial}_{\mu}\psi -e \gamma^{\mu} A_{\mu}\psi 
-m(1+g{\cal G})  \psi =0 . \eqno{ (2.3a)}  $$
Also, we can write the equation of motion for the gravitational field 
$$  \partial_\mu\partial^\mu  {\cal G} =- mg \bar \psi \psi . \eqno{ (2.3b)} $$
Under the static approximation for the gravitational field, we can derive the Dirac 
equation of eq.(1.1) for fermions in the gravitational potential. 

\section{Non-relativistic Gravitational Potential }

\subsection{Foldy-Wouthuysen transformation}
Now, the Hamiltonian of the Dirac equation in the gravitational field can be written as
$$ H=-i \bm{\nabla}\cdot \bm{\alpha}+ \left(m -{GmM\over r}\right) 
\beta  .  \eqno{(3.1)} $$
This Hamiltonian  can be easily reduced to the non-relativistic equation of motion 
by making use of the Foldy-Wouthuysen transformation \cite{bd}. 
Here, we only write the result in terms of the Hamiltonian $H$
$$ H=m+{\bm{p}^2\over 2m}- {GmM\over r} +
{1\over 2m^2}{GmM\over r} \bm{p}^2 
-{1\over 2m^2}{GMm\over r^3} (\bm{s}\cdot \bm{L})  \eqno{(3.2)} $$
where the last term denotes the spin-orbit force, but we do not consider 
it here. Now, we want to make the classical limit to derive the Newton equation. 
In this case, it is safe to assume the factorization ansatz for the third term, 
that is,
$$ \left\langle {1\over 2m^2}{GmM\over r} \bm{p}^2 \right\rangle =
\left\langle{1\over 2m^2}{GmM\over r}\right\rangle 
\left\langle\bm{p}^2\right\rangle .  \eqno{(3.3)}  $$
By making use of the Virial theorem for the gravitational potential
$$ \left\langle {\bm{p}^2\over m} \right\rangle =\left\langle {GmM\over r} \right\rangle 
 \eqno{(3.4)}  $$
we obtain the new gravitational potential for the Newton equation
$$ V(r)= - {GmM\over r} +{1\over 2mc^2}\left({GmM\over r}\right)^2  \eqno{(3.5)}  $$
where we explicitly write the light velocity $c$ in the last term of the equation. 

\subsection{Large Component Treatment}
The new gravitational potential can also be obtained from the equation for the large 
component in the Dirac equation. By denoting the wave function $\psi$ as 
$ \displaystyle{ \psi =\pmatrix{ \phi \cr \chi \cr } }$, 
we can rewrite the Dirac equation 
$$ \left( m-{GmM\over r}-E \right) \phi +\bm{\sigma}\cdot \bm{p}\chi =0, \ \ \ \ \ $$
$$\bm{\sigma}\cdot \bm{p} \phi - \left(E+ m-{GmM\over r} \right) \chi  =0 . 
 \eqno{(3.6)} $$
By denoting $E=m+E_{nr}$, we obtain the equation for $\phi$ 
$$ \left[ {\bm{p}^2\over 2m}- {GmM\over r} +{1\over 2m}\left({GmM\over r}\right)^2  
-{1\over 2m^2}{GMm\over r^3} (\bm{s}\cdot \bm{L}) \right. $$
$$\left.  -{(E_{nr}^{(0)})^2\over 2m} \right] \phi =E_{nr} \phi \eqno{(3.7)} $$ 
where, in the last term, we replace $E_{nr}$ by $E_{nr}^{(0)}$ which is the eigenvalue 
of the unperturbed Hamiltonian. In this case, we obtain the total 
gravitational potential 
$$ V(r)=- {GmM\over r} +{1\over 2mc^2}\left({GmM\over r}\right)^2  $$ 
which is just the same as eq.(3.5). As mentioned above, Coulomb potential has 
no such effect from the non-relativistic reduction, and 
therefore, this type of the additional potential is special for the scalar 
potential.

\section{Angular Velocity Shift of Mercury and GPS Satellite }
The Newton equation with the new gravitational potential can be written as
$$ m \ddot{r} = -{GmM\over r^2} +{\ell^2\over mr^3} +{G^2M^2m\over c^2r^3} . 
 \eqno{(4.1)}  $$
Therefore, we can introduce a new angular momentum $L$ as
$$ L^2=\ell^2 +{G^2M^2m^2\over c^2}.  \eqno{(4.2)}  $$
Further, we define the angular velocity $\omega$ and radius $R$ by
$$ \omega \equiv {\ell\over mR^2}, \ \ \ \ 
R\equiv {\ell^2\over GMm^2(1-\varepsilon^2)^{3\over 4}}   \eqno{(4.3a)}  $$
where $\varepsilon$ denotes the eccentricity. Correspondingly, we can define a new 
angular velocity $\Omega$ associated with $\omega $ as
$$ \Omega^2 \equiv \omega^2 +{G^2M^2\over c^2R^4}=\omega^2 
\left(1+ {G^2M^2\over c^2R^4 \omega^2}
\right) \equiv \omega^2(1+\eta)  \eqno{(4.3b)}  $$
where $\eta$ is defined as
$$ \eta =  {G^2M^2\over c^2R^4 \omega^2} .  \eqno{(4.4)}  $$
The equation (4.1) can be immediately solved, and one finds the solution of the orbit 
$$ r={A\over{1+ \varepsilon \cos \left( {L\over \ell}\varphi \right) }}  \eqno{(4.5)}  $$
where $A$ and $\varepsilon$ are given as
$$ A={L^2\over GMm^2 }, \ \ \ \ \varepsilon=\sqrt{1+{2L^2E\over m(GmM)^2}}. 
 \eqno{(4.6)}  $$
Physical observables can be obtained by integrating 
$\dot{\varphi}= {\ell\over mr^2}$ over the period $T$
$$ {\ell\over m}\int_0^T dt = \int_0^{2\pi} r^2 d\varphi 
= A^2  \int_0^{2\pi} {1\over{\left(1+ \varepsilon \cos 
\left( {L\over \ell}\varphi \right)  \right)^2}} d\varphi.  \eqno{(4.7)}  $$
This can be easily calculated to be 
$$ \omega T=2\pi(1+2\eta)\left( 1-\varepsilon\eta \right)
\simeq 2\pi\{ 1+(2-\varepsilon) \eta\}  \eqno{(4.8)}  $$
where $\varepsilon$ is assumed to be small. 
Therefore, the new gravity potential gives rise to the advance shift 
of the angular velocity, and it can be written as 
$$ \left({\Delta \omega\over \omega}\right)_{th} \simeq (2-\varepsilon)\eta . 
 \eqno{(4.9a)}  $$
This is a physical observable which indeed can be compared to experiment. 
Here, it may be interesting to comment on eq.(4.8). It shows that one may also 
discuss the physics of the deviation in terms of the time shift. 
$$ \left({\Delta T\over T}\right)_{th} \simeq (2-\varepsilon)\eta . \eqno{(4.9b)}  $$
In fact, the shift of the GPS satellite is often described in terms of the time shift. 

\subsection{Mercury Perihelion Shift }
The Mercury perihelion advance shift $\Delta \theta $ is well known to be \cite{misner}
$$ \Delta \theta \simeq 42\  '' \ \ {\rm per \ \  100 \ year } . \eqno{(4.10)}  $$
Since Mercury has the $0.24$ year period, it can amount to the shift ratio 
$\delta \theta$ 
$$ \delta \theta_{obs} \equiv \left({\Delta \omega\over \omega}\right)_{obs} 
\simeq 7.8 \times 10^{-8}.  \eqno{(4.11)}  $$
The present theoretical calculation shows 
$$ \eta =  {G^2M^2\over c^2R^4 \omega^2} \simeq  2.65 \times 10^{-8}   \eqno{(4.12)} $$
where the following values are used for the Mercury case
$$ R=5.73 \times 10^{10} \ {\rm m}, \ \ \  M= 1.989 \times 10^{30} \ {\rm kg}, 
\ \ \ \omega=8.30 \times 10^{-7}.  $$
Therefore, the theoretical shift ratio $ \delta \theta_{th} $ becomes 
$$  \delta \theta_{th} \equiv \left({\Delta \omega\over \omega}\right)_{th} 
\simeq 4.8 \times 10^{-8}  \eqno{(4.13)}  $$
which should be compared to the value in eq.(4.11). As can be seen, this agreement 
is indeed remarkable since there is no free parameter in our theoretical 
calculation. 

\subsection{GPS Satellite $\omega-$Shift }
Many GPS satellites which are orbiting around the earth should be influenced  
rather heavily by the new gravitational potential. The GPS satellite $\omega-$shift 
can be estimated just in the same way as above, and we obtain
$$ \eta = {G^2M^2\over c^2R^4 \omega^2} \simeq 1.69 \times 10^{-10}  \eqno{(4.14)}  $$
where we employ the following values for the GPS satellite   \cite{park,nere,bah}
$$ R=2.6561 \times 10^{7} \ {\rm m}, \ \ \  M= 5.974 \times 10^{24} \ {\rm kg}, $$
$$\ \ \ \omega=1.4544 \times 10^{-4}  
 \eqno{(4.15)}  $$
since the satellite circulates twice per day. 
Therefore, the advance shift of the GPS satellite becomes
$$ \left({\Delta \omega\over \omega}\right)_{th} \simeq  3.4 \times 10^{-10}. 
 \eqno{(4.16)}  $$
This should be compared to the observed 
value of 
$$ \left({\Delta \omega\over \omega}\right)_{exp} \simeq  4.5 \times 10^{-10} . 
 \eqno{(4.17)}  $$
As seen from the comparison between our calculation and the observed value, 
the present theory can indeed achieve a remarkable agreement with experiment.

\section{Prediction from General Relativity }
Here, we briefly discuss the calculated result by the general relativity. 
For the Mercury perihelion shift, the result is quite well known, and it can be 
written in terms of the angular shift. In fact, the angular variable $\varphi$ 
is modified by the general relativity to 
$$ \cos \varphi \longrightarrow \cos (1-\gamma) \varphi   \eqno{(5.1)} $$
where $\gamma$ is found to be 
$$ \gamma = {3G^2M^2\over c^2R^4 \omega^2} . \eqno{(5.2)}  $$
This change of the shift in the angular variable explained 
the observed Mercury perihelion shift. However, as can be seen from eq.(4.5), 
this effect vanishes to zero in the case of $\varepsilon =0$, that is, 
for the circular orbit. This is, of course, unphysical in that the effect of 
the general relativity is valid only for the elliptic orbit case. 
In Newton dynamics, the angular momentum $\ell$ is the only quantity 
which can be affected from the external effects like the general relativity 
or the additional potential. As can be seen from eq.(4.5), the effect of 
the angular variable in eq.(5.1) induces the change in $A$ of eq.(4.5) as well, 
and this is indeed inevitable. Therefore, we should calculate the physical 
observables as to how the general relativity can induce the perihelion shift, and 
we find
$$ \omega T \simeq 2\pi\{ 1-2(2-\varepsilon) \gamma \} . \eqno{(5.3)}  $$
Unfortunately, this is not the advance shift of the Mercury perihelion, and 
instead, it is a retreat shift. 

This is a reflection of the important consequence in physics that one should 
evaluate physical observables in order to compare theory to experiment. 
In this respect, the angular shift calculated by the general relativity 
\cite{ein,misner} is not a physical observable. In fact, as can be easily seen, 
the angular shift at one fixed point of the perihelion cannot take into account 
the effects that should arise from all the orbital trajectories of the Mercury 
since it is rotating around the sun. The shift must be the result of the 
integrated quantity over the orbit, and it is not a quantity just at one point of 
the trajectory. Since the observed shift of the Mercury perihelion was obtained 
in comparison with the one that was made some years before, there is no chance that 
the observed perihelion shift is directly related to the angular shift in eq.(5.1).


\end{document}